\documentclass[conference]{IEEEtran}
\IEEEoverridecommandlockouts

\usepackage{code}
\usepackage{graphicx}
\usepackage{balance}
\usepackage{multirow}
\usepackage{multicol}
\usepackage{listings}
\usepackage{booktabs}
\usepackage{subfig}
\usepackage{url}
\usepackage{comment}
\usepackage{xcolor}
\usepackage{xspace}

\definecolor{dkgreen}{rgb}{0,0.5,0}
\definecolor{dkred}{rgb}{0.5,0,0}
\definecolor{gray}{rgb}{0.5,0.5,0.5}

\lstdefinestyle{javastyle} {
language=Java,
basicstyle=\ttfamily\bfseries\footnotesize,
  morekeywords={virtualinvoke},
  keywordstyle=\color{blue},
  ndkeywordstyle=\color{red},
  commentstyle=\color{dkred},
  stringstyle=\color{dkgreen},
  numbers=left,
  breaklines=true,
  numberstyle=\ttfamily\footnotesize\color{gray},
  stepnumber=1,
  numbersep=10pt,
  backgroundcolor=\color{white},
  tabsize=4,
  showspaces=false,
  showstringspaces=false,
  xleftmargin=.23in
}
\begin{document}

\title{Mind the Gap: The Difference Between Coverage and
  Mutation Score Can Guide Testing Efforts}

\author{
\IEEEauthorblockN{Kush Jain}
\IEEEauthorblockA{\textit{Carnegie Mellon University} \\
Pittsburgh, Pennsylvania}
\and

\IEEEauthorblockN{Goutamkumar Tulajappa Kalburgi}
\IEEEauthorblockA{\textit{Northern Arizona University} \\
Flagstaff, Arizona}
\and

\IEEEauthorblockN{Claire Le Goues}
\IEEEauthorblockA{\textit{Carnegie Mellon University} \\
Pittsburgh, Pennsylvania}
\and
\IEEEauthorblockN{Alex Groce}
\IEEEauthorblockA{\textit{Northern Arizona University} \\
Flagstaff, Arizona
}}


\newcommand{\mr}[2]{\multirow{#1}{*}{#2}}
\newcommand{\mc}[3]{\multicolumn{#1}{#2}{#3}}
\newcommand{\um}{\texttt{universalmutator}\xspace}
\newcommand{\clg}[1]{\textcolor{blue}{#1}}
\newcommand{\adg}[1]{\textcolor{purple}{#1}}
\newcommand{\kj}[1]{\textcolor{olive}{#1}}

\newcommand{\averageprojvariance}{402}
\newcommand{\averagevariance}{604}
\newcommand{\outliertotalfiles}{26}
\newcommand{\outliertestissues}{12}
\newcommand{\outlierumissues}{7}
\newcommand{\outlierunclear}{7}
\newcommand{\allcorr}{0.7479}
\newcommand{\covcorr}{0.2066}
\newcommand{\allrsquared}{0.573}
\newcommand{\allr}{0.757}
\newcommand{\covrsquared}{0.021}
\newcommand{\covr}{0.145}

\maketitle

\begin{abstract}
An ``adequate'' test suite should effectively find all
inconsistencies between a system's requirements/specifications and its
implementation. Practitioners frequently use code coverage to
approximate adequacy, while academics argue that mutation score may
better approximate true (oracular) adequacy coverage. High code
coverage is increasingly attainable even on large systems via
automatic test generation, including fuzzing. In light of all of these
options for measuring and improving testing effort, how should a QA
engineer spend their time? We propose a new framework for reasoning
about the extent, limits, and nature of a given testing effort based
on an idea we call the \emph{oracle gap}, or \emph{the difference
between source code coverage and mutation score for a given software
element}. We conduct (1) a large-scale observational study of the
oracle gap across popular Maven projects, (2) a study that varies
testing and oracle quality across several of those projects and (3) a
small-scale observational study of highly critical, well-tested code
across comparable blockchain projects. We show that the oracle gap
surfaces important information about the extent and quality of a test
effort beyond either adequacy metric alone. In particular, it provides
a way for practitioners to identify source files where it is likely a
\emph{weak} oracle tests \emph{important} code.
\end{abstract}




\section{Introduction}

Which cryptocurrency project has better testing practices, Bitcoin Core or Algorand?
Both projects' tests cover their critical core transaction verification logic
reasonably well: Algorand's tests achieve 90\% statement coverage on the files
in question, while Bitcoin's cover an astonishing 98.7\% 
of the statements in equivalent core code. Using this (very common, in
practice~\cite{Discontents,codeCoverageGoogle}) test suite adequacy metric, although Algorand is
certainly well-tested, Bitcoin has the edge.  Of course, the question
being asked is not specific to cryptocurrency codebases, but can more
generally be asked of any two testing efforts, including different
tests for the same project, or the same project at different times.

It is well known that \emph{executing} code and actually fully
functionally testing it are not the same thing. Code coverage is informative,
but is really a one-way test rather than a true measure of
test suite quality: low coverage is bad, but high coverage does not mean
a test suite is truly adequate. Another long-studied
way to measure adequacy relies on \emph{mutation analysis}, which checks how
well tests detect syntactic changes injected
into source code files. Mutation analysis \emph{incidentally} measures
code coverage (you cannot detect changes you do not run) but
\emph{fundamentally} measures oracle quality/power:
the ability to tell ``good'' from ``bad'' code~\cite{StaatsOracle}.

Returning to our cryptocurrency project example, armed with this
expensive but arguably more precise metric we can again ask: who
is winning, Algorand or Bitcoin? Using an any-language mutation
testing framework~\cite{regexpMut} on Bitcoin's 
transaction verification code we find a more than respectable 75.85\% mutation score. Meanwhile,
using comparable settings, Algorand's tests boast a remarkable
99.7\% mutation score! Perhaps Algorand is ``winning'' after all? Modern test
adequacy research focuses on correlating either coverage, mutation score, or
both, with the prevalence of faults, claiming e.g., that coverage
is ``good enough'' or that mutation's expensive ``more informative number'' is
needed~\cite{whatdoweknow,covdev,ThierryStudy}.  Here, though, mutation score
is not a ``refinement'' of coverage, but yields a substantially different
assessment of effectiveness.  

In this paper, we present evidence in favor of a new way of thinking about these
metrics, and in particular how to use the relationship between them to inform
testing efforts. Our opening question is a red herring: in practice, which of
two different projects has better tests is not nearly as important a question as
``how should Bitcoin's (or Algorand's) test engineers spend their time?'' Neither coverage nor
mutation score alone explains \emph{where a test effort has been effective thus
  far, and where effort should now be directed.}

The structure of testing advice to date,
regardless of metric, is most easily summed up as 
``write more tests.''  As automated testing grows, however, improving test
coverage and actual \emph{oracle} power
often diverge into separate engineering efforts and 
research topics. For example, changing
fuzzers or using symbolic execution to cover additional obscure paths is a poor
use of time if assertion weakness is the primary limit of a given test effort. Meanwhile, adding
new checks is of limited value when a testing effort is not able to
find most outright \emph{crashes}. The naive approach, to
focus on improving coverage until it is nebulously ``high'' before working on
improving assertions, ignores the fact that in many code bases, once the most critical
functionality is covered, there may be more utility in improving the existing tests
than in ``racking up'' coverage gains by simply running uninteresting code.

\looseness-1
We define the \emph{oracle gap} as \emph{the difference between source code
  coverage and mutation score for a given software element}. Naive,
automatically-generated tests that cover code without strong assertions
straightforwardly produce positive gaps; basic coverage is easier to achieve
than high mutation scores. Conversely, code that is well tested but not 
well covered produces negative gaps: coverage is low, but the mutation score is
high because the provided oracles for covered code are strong, and the
covered/tested elements contribute a large portion of the mutants
(e.g., where complex arithmetic and logical code with many mutants is
well tested, but surrounding initialization or logging calls, where
only statement deletion applies, are
poorly tested).


\looseness-1
Previous efforts to establish test adequacy
measures tend to focus on a single
number corresponding to ``testedness''~\cite{testedness};  monolithic scores
will always struggle to
distinguish very different kinds of test effort and are thus difficult to \emph{act
on} cost-effectively.  Making such metrics actionable is the goal of this paper,
and in our experiments we show that the oracle gap is more useful for this
purpose than either metric alone, or alternatives like 
pseudotesting~\cite{pseudotestingorig,
  pseudotestingstudy, descartes}.
In many real-world projects, the code that \emph{is covered} is
well-tested, perhaps because finite resources are best devoted to code where the
impact of bugs is higher, but seldom-used or merely ``cosmetic'' code
is not executed by tests.  It is not obvious
that developers are doing the wrong thing, in such cases. 
However, it is still helpful for developers to \emph{know}
that that's what their test effort looks like, to help allocate finite
QA resources.

In particular, the primary use of the oracle gap is to
identify code (files or even functions) with unusually high coverage relative to
mutation score.  In some, obvious instances, these may be logging or
other code that is easy to execute and unimportant to test.  However, such
unusual \emph{positive oracle gaps} should be prioritized for
examination by developers or test engineers, in that they likely
represent cases where a portion of code considered important (hence
the high coverage) has not been sufficiently tested.  Such cases may
arise due to lack of specification effort, or due to bugs in testing
(e.g., an assertion with an implication that never holds, an
apparently useful but in practice vacuous check).  Previous
approaches to similar problems do not address these problems. 
Checked coverage~\cite{ZellerCheckedCov} and related methods~\cite{GapStruct}, for example,
cannot detect cases where a statement is covered and flows to an
assertion, but the assertion is written in such a way that the
outcome is never false.  Prioritizing code \emph{to examine for oracle
weaknesses} by code coverage alone is obviously useless.
Prioritizing by mutation score alone similarly will usually simply
point out poorly-covered (and often unimportant) code.  We therefore
propose a simple, primary use for the oracle gap: \emph{examine source
  elements (usually files) exhibiting unusual positive oracle gaps, as
  these are the likely locations of missing or buggy assertions.}

%

Our contributions are of four kinds.  First, fundamentally, we define
the \emph{oracle gap} and the
   \emph{covered oracle gap} to measure the tendency of a test effort
  to 1) favor executing code over checking its behavior for
  correctness, 2) favor checking correctness of executed code over
  covering structural code elements or 3) balance these goals.
Second, we gather empirical data from real-world Java projects,
  investigating the distribution of raw and
  covered oracle gap, showing that knowing the oracle gap adds useful information
  over either coverage or mutation score alone.
Third, a synthetic examination
  of oracle gap on a subset of Java projects demonstrates the oracle gap's ability to identify gaps in testing
  efforts or, put differently, to provide accurate, actionable testing advice. 
Finally, we present a qualitative investigation of how oracle gap varies across equivalent
  implementations from several
  comparable projects that should be thoroughly tested, and thus surfaces differences in test approach and quality not otherwise visible.


\section{The Oracle Gap}

The most common
approximation for test adequacy in practical use is \emph{code coverage}, or the
percentage of the code base (measured in terms of lines or branches, typically)
the test suite executes.  Although cheap to compute, it measures
execution, not testedness.  \emph{Mutation analysis}  checks how
well a project's tests detect syntactic changes injected
into project source code~\cite{HintsOnTestDataSelection}. Mutation analysis incidentally measures
code coverage, but more fundamentally evaluates the test suite's ability to
detect bad code.

Mutation analysis is well established in the academic
literature~\cite{demillo1978hints, budd1979mutation, jia2008constructing, zhangPMT}.  As a practice, however, it has only recently begun to achieve 
industrial penetration~\cite{PetrovicMutationGoogle,BellerFacebookMutation}, in large part due to its
computational expense, and seldom as an actual adequacy metric (vs. a
pinpoint for test issues in \emph{newly committed code}).  
Prior work has extensively examined the correlation between various mutants generated by mutation analysis 
and real world faults in code, finding that in many cases mutation analysis can mimic such faults 
in code \cite{JustMutationFault}. As a result, researchers recommend using
mutation score to measure adequacy in terms of a test suite's likelihood to
detect bugs. 
In contrast, defenses of code coverage often amount to
claims that, in practice, it is predictive enough of mutation score
that it is reasonable to use it in its place.

We argue in favor of reasoning about coverage and mutation score in
tandem to inform both assessments of current testing efforts, and
decisions about where to expend future effort.  Current practice,
focusing on mutants of ``new code,'' fails especially when future
effort should re-visit older, but weak, tests.  We define the \emph{oracle gap}
for a given test effort as:

\begin{equation}
\mathtt{oracle\_gap}_{X,T} = cov(X,T) - mut(X,T)
\end{equation} 

Where $X$ corresponds to the unit of analysis (file, project, etc) and $T$ to
its test suite.  Computing the difference between 
two different measures (i.e., statements covered and
mutants killed) that are themselves measured as fractions of different units may at first appear improper; however, we could
formally map any two metrics into an abstract adequacy measure, using
fixed coefficients of 1.0, or by defining a suitable transfer
function.  Informally, the key is that the difference is only
interesting in terms of sign and approximate magnitude.  In practice,
a comparison of whether code is ``more covered'' or ``more checked''
is straightforward enough for conception and application.  In the
research literature, informal comparisons of, e.g., branch and statement
coverage are not infrequent \cite{covComparisons,
  mutationStatementBranchComparison}.  Indeed, PIT and other mutation
tools arguably report a limited kind of ``binary'' gap, by noting (only)
individual lines of code that are covered, but whose mutations are not
detected.  In a sense we simply extend this notion to files and other
larger coverage units including entire projects, with a finer
granularity for mutation score.
Our first
research question is therefore:

\begin{quote}
\textbf{RQ1}: What is the empirical relationship between mutation score and coverage? 
\end{quote}

We study this question on a large dataset of Java projects
(Section~\ref{sec:javastudy}).  We find that mutation score and coverage are indeed
positively correlated (as shown in many previous studies \cite{papadakis2018mutation, inozemtsevaCoverageTestEffectiveness, testedness}),
although with high variance.  
Despite this, there should not be (and in fact, we do not find) a
perfect correlation; otherwise,
it would never make sense to use the computationally more expensive
mutation score. Instead, we find that the two correlate
particularly well at low coverage values (where mutation score rarely exceeds
file coverage).   Better covered files show significantly more variation, 
with frequent large positive gaps between coverage and mutation
score---indicating code that is executed, but poorly checked.

On the face of it, then, divergences between coverage and mutation score might
be most informative when the implicit measure of coverage contained in
mutation score is removed. We therefore distinguish
between raw oracle gap and \emph{covered oracle gap}, or
the gap between code coverage and mutation score
\emph{over covered code only}. Prior work~\cite{PapadakisStudy,ThierryStudy}
runs mutation testing on all code, giving a score strongly related to coverage.
Recent work by
Google~\cite{PetrovicMutationGoogle} 
only mutates covered code, because mutations to uncovered
code are obviously not going to be detected, even if they induce
outright crashes.
This informs our second research question (also Section~\ref{sec:javastudy}):

\begin{quote}
\textbf{RQ2}: How does the correlation between mutation score and coverage
differ when only considering covered code (lines)? 
\end{quote}

Covered oracle gap is a more novel
point of view on a testing effort.
The covered oracle gap directly speaks to the quality of the developer-written oracles,
taking into account that one cannot detect
mutations of unexecuted code.

Both types of oracle gap speak to the
quality and form of a testing effort.  However, our first two questions are
fundamentally descriptive, retroactively using the oracle gap to evaluate
testing efforts post facto.  For the oracle gap to be useful, it must be better
able to precisely identify actionable inadequacies in testedness. 
Therefore,
we ask the question:

\begin{quote}
\textbf{RQ3}: Can oracle gap clarify testing \emph{problems}?
\end{quote}

We answer this question via a synthetic experiment where we artificially
vary the coverage and oracle power of well tested real-world Java
code (Section~\ref{sec:synthetic}).  We find that the oracle gap does ``move''
as expected, increasing as assertions are removed or coverage is added,
decreasing otherwise, corresponding to expected advice to focus attention on
either coverage or test oracle power, respectively. We moreover find that it
does so more precisely than alternative metrics. 

%
%

Finally, we look at a smaller dataset of comparable programs where we
could identify central functionality that \emph{should} be well tested to
answer:

\begin{quote}
\textbf{RQ4}: What are the implications (and causes) of a small or
large, and positive or negative, oracle gap, across comparable real-world test
efforts? 
\end{quote}

To answer this question, we performed a case study of high market cap
cryptocurrencies' transaction and block validation code (Section~\ref{sec:crypto}).  We manually examined
test suites and determined the reason(s) for different oracle gaps.

\section{Oracle gap on large Java projects}
\label{sec:javastudy}

\begin{table}
\caption{\small Descriptive Statistics for Large Java Corpus.}
\centering
\begin{tabular}{lrrr|r}
\toprule
\bf Statistic                   & \bf Mean        & \bf Median       & \bf Std. Dev.            & \bf Total    \\
\midrule
\mc{5}{c}{\emph{Per project}}  \\
\midrule
\#files                 & 267.8           & 156.0              & 447.0                    & 12051.0        \\
mutation analysis     & \mr{2}{51.8}            & \mr{2}{30.3}             & \mr{2}{236.6}                    & \mr{2}{54758.9}    \\
 runtime (mins)       &                 &         &                        &                      \\
lines of code         & 10298.2         & 5767.0             & 11715.0                  & 463417.0       \\
\# tests                & 1277.6          & 483.0              & 1881.2                   & 42161.0        \\
\midrule
\mc{5}{c}{\emph{Per file}} \\
\midrule
line coverage              & 63.8            & 81.3             & 39.5                     & -            \\
\#mutants                  & 41.9            & 30.5             & 36.0                     & 44320.0        \\

\bottomrule
\end{tabular}
\label{tab:corpus}
\end{table}

Our first two research questions ask:

\begin{quote}

\textbf{RQ1} What is the empirical relationship between mutation score
  and coverage?

\vspace{1ex}
\textbf{RQ2} 
How does the correlation between mutation score and coverage
differ when only considering covered lines? 
\end{quote}

We conduct an observational
study of large Java projects, providing a high level view of how mutation score and coverage are related,
as well as general oracle gap trends.

\subsection{Experimental Setup}

\subsubsection{Dataset}
We aimed to construct a dataset of large, well-maintained, popular real-world
Java projects.  
Our analysis requires both coverage computation and mutation analysis, and so we
sought projects that used the Maven build system and provided coverage
reports. We collected Java projects from GitHub
ordered by star count (a proxy for popularity) as well as repositories
from top open source organizations with more than 20 stars. 
Our final corpus contains 45 projects, totalling 463,417 lines of code across 12,051 files.
Table~\ref{tab:corpus} lists descriptive statistics. Although there is significant
variation, the average file is
fairly well covered (63.8\%) by the projects' extant test suites. 


\subsubsection{Setup and Analysis}
We use the \um~\cite{universalMutator}, a multi-lingual regular-expression-based
tool for mutant generation, to compute mutation score. \um's operators
are the usual combination
of arithmetic (e.g., replace {\tt +} with {\tt -}), logical (replace
{\tt \&\&} with {\tt ||}), statement deletion, and control flow (e.g.,
adding {\tt break} or {\tt continue}) changes.  We used
\um instead of the popular PIT tool for Java~\cite{pit}
first because \um can handle C and C++ 
as well as Java, allowing us to use consistent settings and operators 
(modulo a few language-specific operators) across all research questions.
Second, we manually investigated
certain mutants to better understand our results. 
PIT mutates code at the
bytecode/ASM level,  while \um produces
easily interpreted source-level mutants. \um has been validated
against other widely used mutation tools for
Java, C++, Rust, and Python mutation, and has a good combination of
low equivalency rates and diversity of mutation operators.

Computing mutation score on all files in every project in our corpus
is computationally intractable. We therefore sampled up
to 100 mutants per file, for
up to 100 files per project.  
We perform stratified sampling by
statement coverage to select the files for analysis per project, bucketing all files into 0-25\% coverage, 25-50\% coverage,
50-75\% coverage and 75-100\% coverage. We randomly select 25 files in each
bucket for buckets with more than 25 files (and all files otherwise). 
We performed no analysis to remove
equivalent mutants, as our use of mutation scores is focused only on
comparing differences across projects, where we can assume
equivalence rates are sufficiently similar across projects to not substantially impact
our results.
Our corpus contains a median of 236 mutants/file, while 
we sample a median 30.5 mutants/file, giving us 
a 12.9\% sample rate overall. This is in line with prior
work~\cite{GopinathSampleSize} that analyzes mutation score in light of
sampling; they
obtain a 7\% error when sampling less than 5\% of
total mutants.



\subsection{Results}

\begin{figure}
  \centering
  \subfloat[Regression Line Plot\label{fig:alllr_plot}]{{\includegraphics[width=7cm]{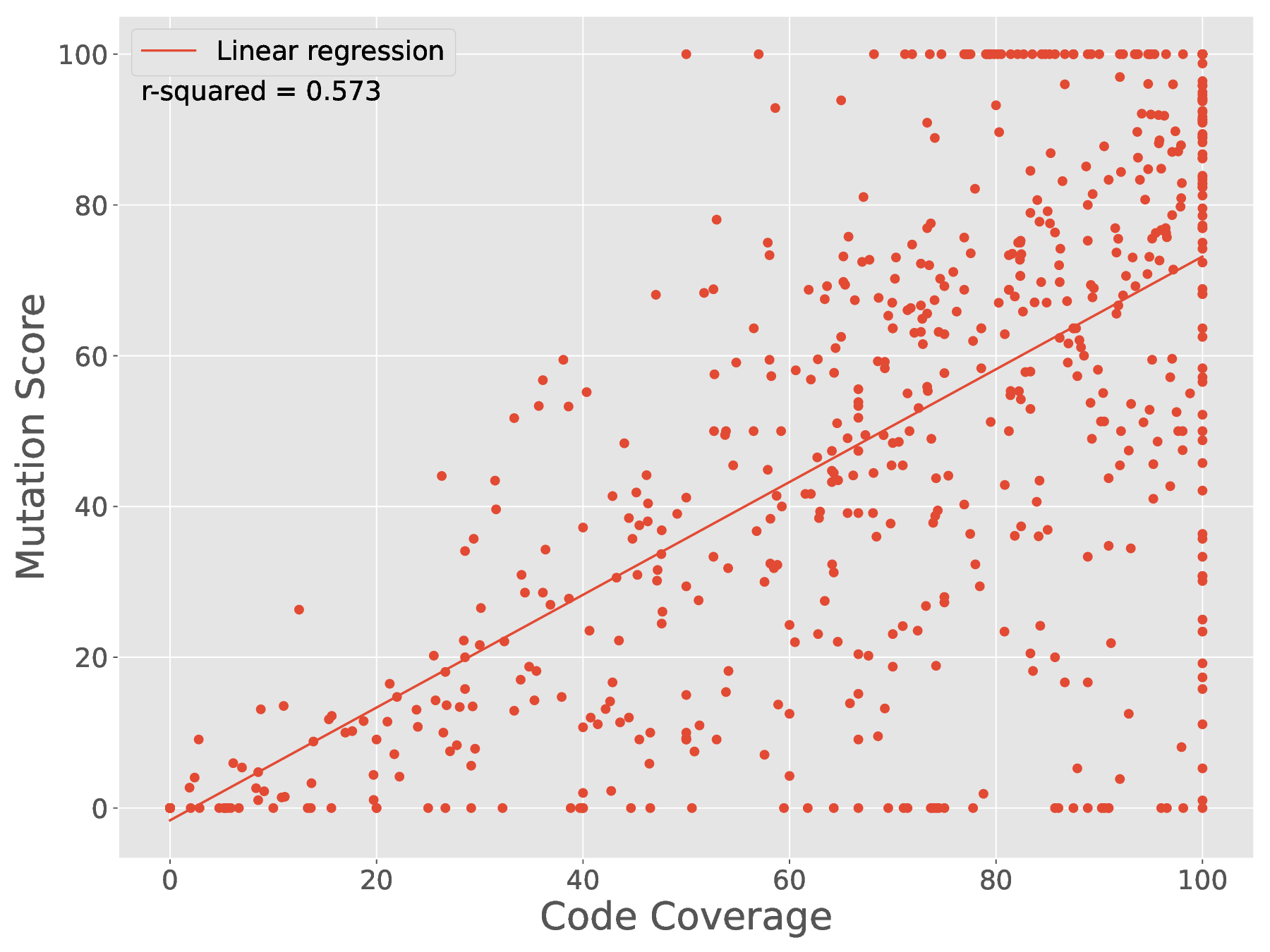} }}%
  \qquad
  \subfloat[Residual Plot\label{fig:alllr_resid}]{{\includegraphics[width=7cm]{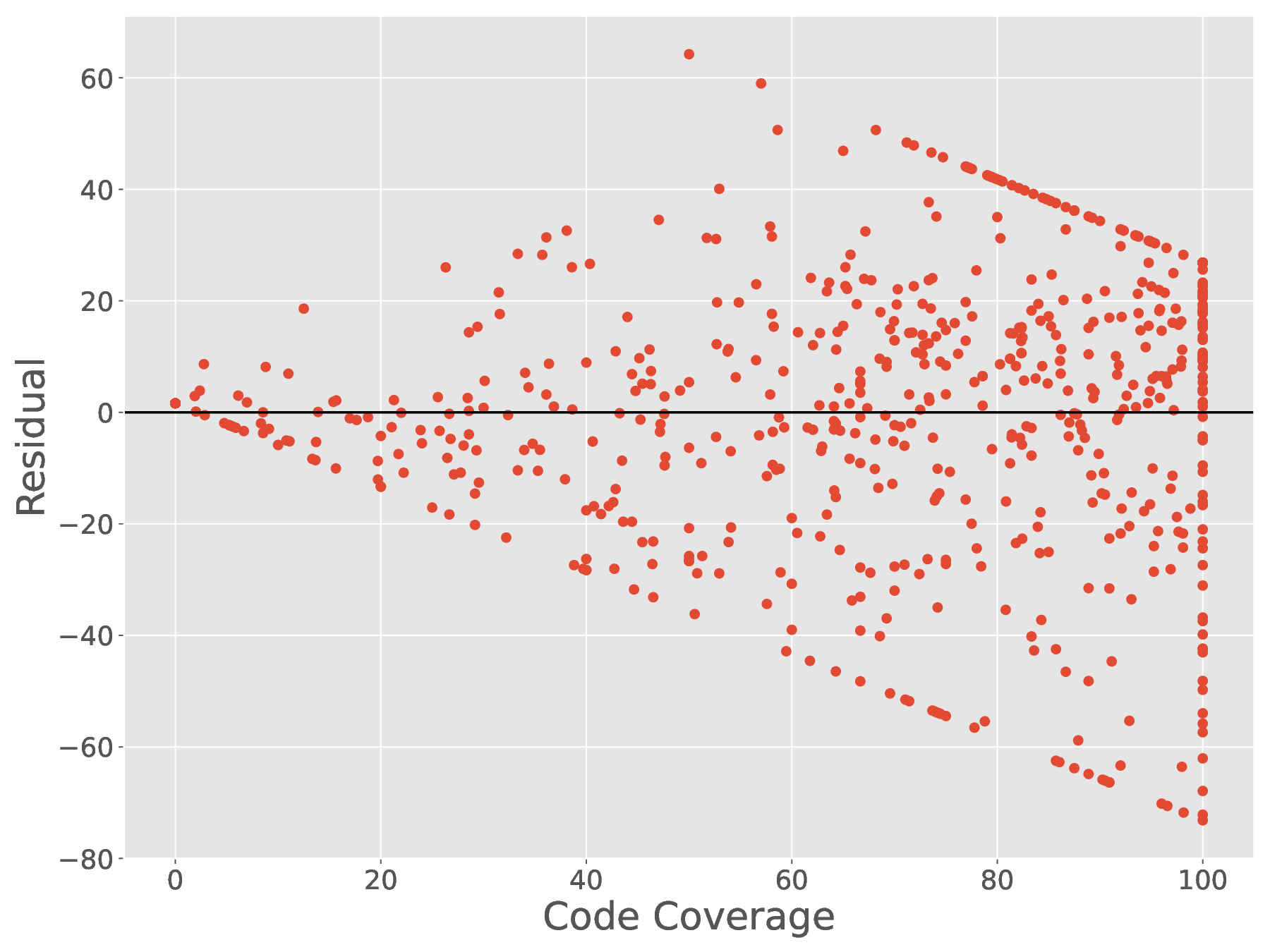} }}%
  \caption{\small Regression and residuals of coverage vs mutation score for all
    lines. Coverage and mutation score are weakly positively correlated.}
  \label{fig:alllr}
\end{figure}

\begin{figure}
  \centering
  \subfloat[Regression Line Plot\label{fig:coveredlr_plot}]{{\includegraphics[width=7cm]{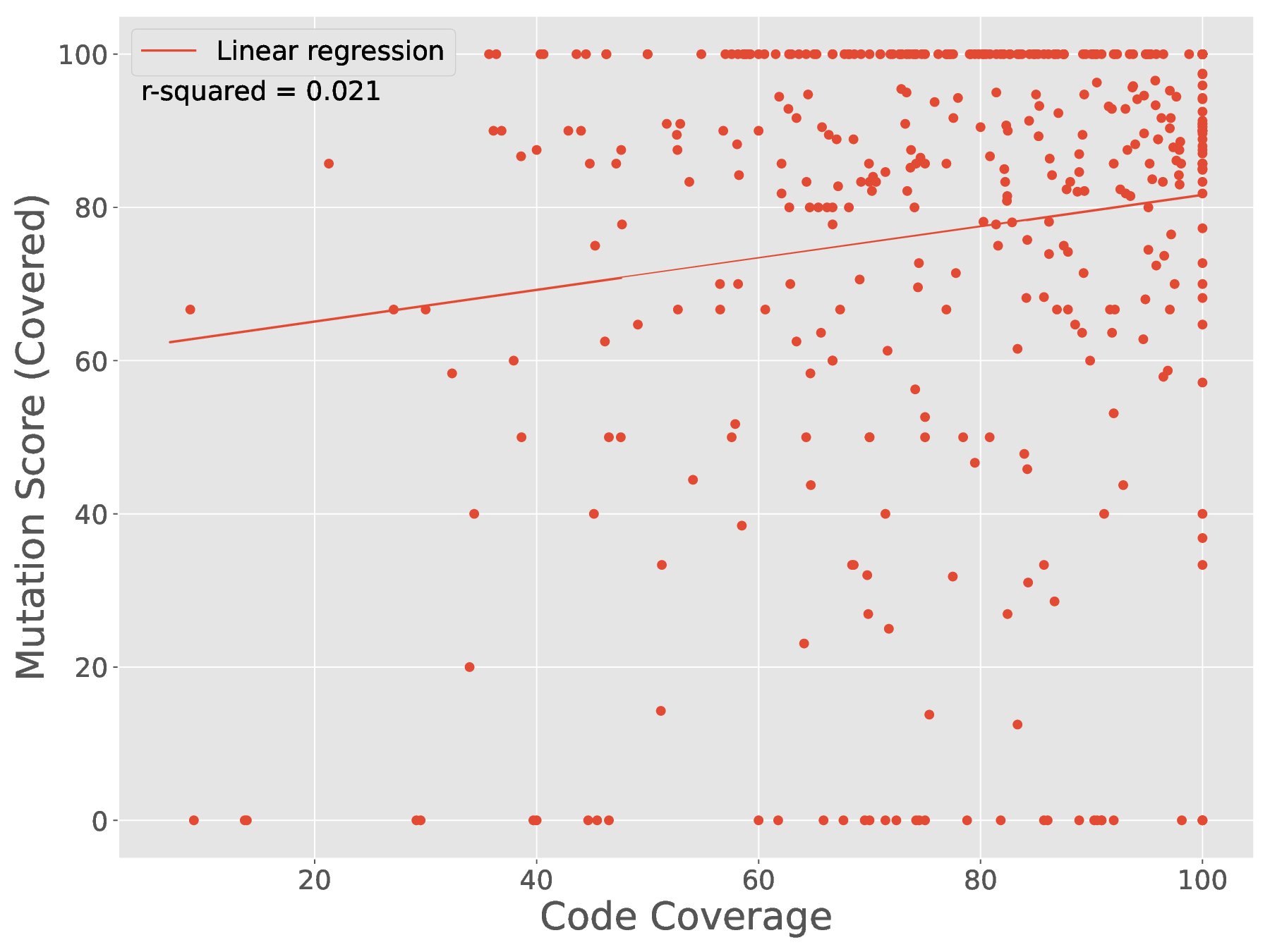} }}%
  \qquad
  \subfloat[Residual Plot\label{fig:coveredlr_resid}]{{\includegraphics[width=7cm]{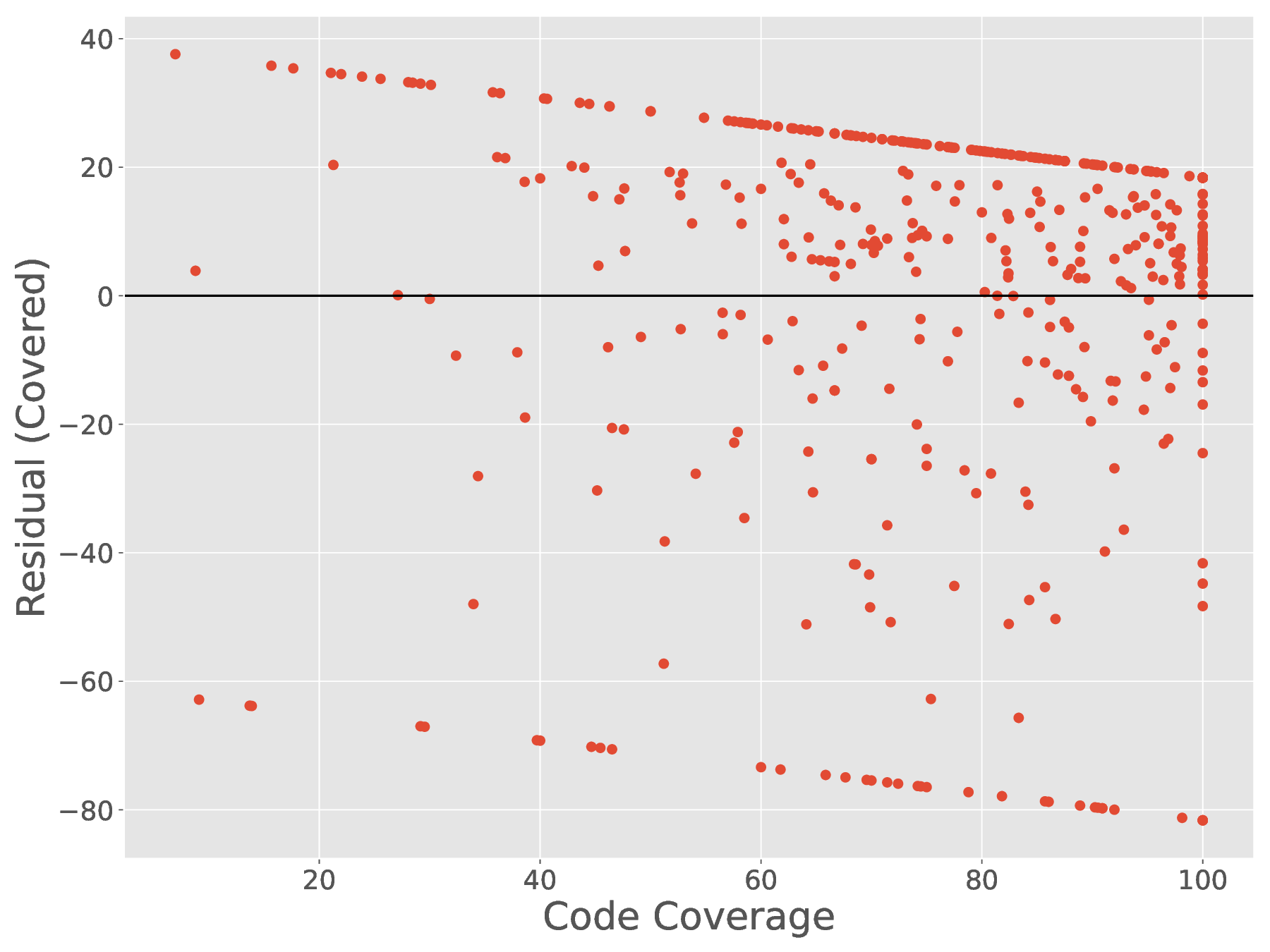} }}%
  \caption{\small Regression and residuals of coverage vs mutation score for covered lines, showing significant noise and no clear correlation.}
  \label{fig:coveredlr}
\end{figure}

Figure~\ref{fig:alllr} shows a linear regression fit
between mutation score over all lines of code and code coverage; Figure
\ref{fig:coveredlr} shows the same over covered lines of
code. Figure~\ref{fig:boxplots} shows
boxplots of raw and covered oracle gaps at different coverage thresholds, and in aggregate.

\subsubsection{Raw Oracle Gap (RQ1)} 
Figure~\ref{fig:alllr} shows a weak positive correlation between mutation score and coverage. 
The correlation coefficient of the regression line is 
{\allcorr}, with an r-value of {\allr}. Variation is moderately explained by the line,
with an $R^2$ value of {\allrsquared}. For low coverage files,
mutation scores tend to be very low, with only a couple of outliers.
Figure~\ref{fig:boxplots} plots oracle gap by coverage range, showing substantial variance (603.92) and
a Pearson correlation
coefficient of 0.4591 between coverage and raw oracle gap. 
Importantly, the residual plot indicates that there is likely \emph{not} a
strictly linear relationship between mutation score and coverage, as residual
magnitude increases with coverage. Instead, the plot suggests that at lower
levels of coverage, the relationship is somewhat linear, but as coverage
increases it becomes a much weaker predictor of mutation score. This is also
visible in Figure~\ref{fig:boxplots}. 

This increased variance at higher
coverage values (for files in the 75\%--100\% coverage bucket, variance is 823.31, compared to a variance of 25.35 for 
files in the 0\%--25\% bucket) suggests that mutation testing and oracle gap are more
informative at higher coverage levels (motivating mutation analysis only once
a certain amount of test effort has been expended).

\subsubsection{Covered Oracle Gap (RQ2)} 
Unlike the relationship between overall mutation score and coverage, the relationship between covered lines and
mutation score \emph{over covered lines only} is much noisier overall, with weaker correlation.  The regression in Figure~\ref{fig:coveredlr} shows a correlation
coefficient of {\covcorr} and an r-value of {\covr}. The ${R}^2$ of {\covrsquared} indicates that very little of the
variation can be explained by the regression line.

\begin{figure}
  \centering
  \includegraphics[width=7cm]{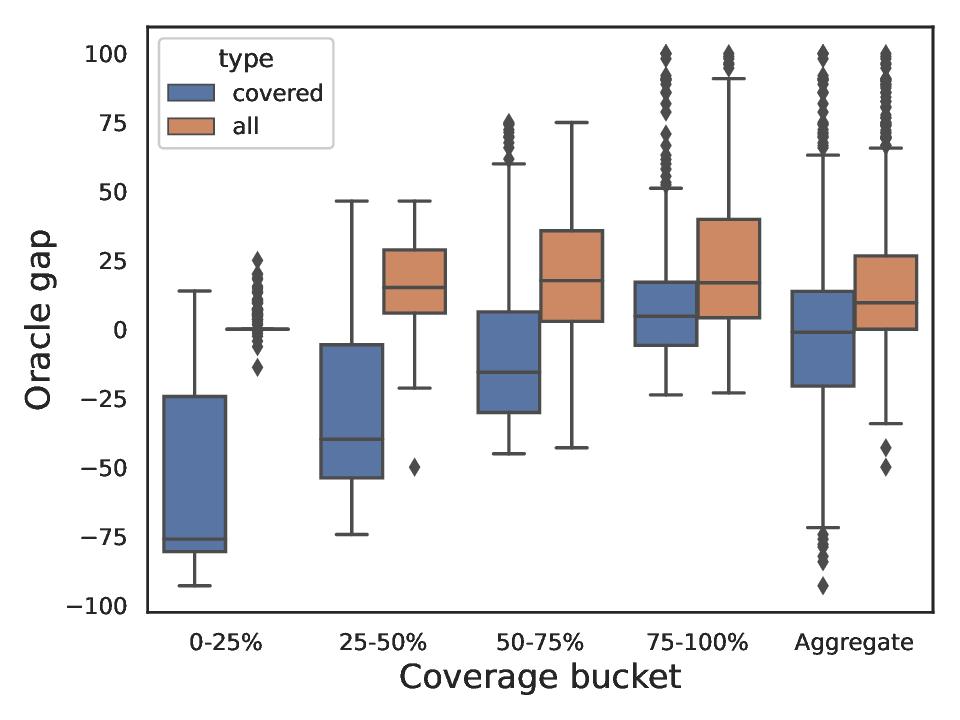}
  \caption{\small Boxplot of oracle gaps, both raw and covered, for the Java dataset.
    Covered oracle gaps tend to increase with coverage, but variance decreases;
    there is high variance overall, though it is higher in aggregate for covered gaps.\label{fig:boxplots}}
\end{figure}

The residual plot, however, depicts an interesting phenomenon: at lower
coverages the relationship tends towards a negative
oracle gap.  This suggest that for poorly-covered files, the outliers tend to be
cases where few lines are being tested, but in general these lines  
are well tested. At higher coverage, the oracle gap trends
towards positive oracle gaps, as would usually be expected given that
code is easier to execute than to check.
Figure~\ref{fig:boxplots} also supports this conclusion, with a slight upwards trend in covered oracle gaps,
as coverage increases (correlation coefficient of 0.5661) and an increased aggregate variance of 1142.74. Additionally, and by
contrast with the results for raw oracle gap, variance in covered gaps
\emph{decreases} at higher coverage
levels (for files with 75\%--100\%, coverage, variance is 742.30 compared to a variance of 1655.82 for 
files with 0\%--25\% coverage).  That is, it appears that better-tested files
tend towards more consistent testing behavior (trending slightly positive in
terms of gap). We examine these phenomena further in Section~\ref{sec:crypto},
on a different, more directly comparable dataset of projects.

To validate our observations, we manually examined the  {\outliertotalfiles} files in our corpus with
greater than 80\% coverage and less than 20\% mutation score. Of these files, at least
{\outliertestissues} clearly had multiple missing
assert statements in their tests, failing to catch obvious defects
introduced by ``gross'' mutation.  Another {\outlierunclear} files had one specific
type of missing assertion, corresponding to a specific mutation operator. Finally, the remaining {\outlierumissues} files were false
positives caused by, e.g., message string or log statement mutations, 
leading to spurious low
mutation scores.

\subsubsection{Discussion}

Based on these results, we observe:
\begin{itemize}
    \item As coverage increases, oracle gap variance also increases, suggesting
      more value in running mutation analysis for test suites that achieve
      higher coverage.
    \item The \emph{covered oracle gap} exhibits significant variance, in
      aggregate, though it decreases substantially as coverage increases. 
      Moreover, while the basic limit that
      low coverage tends to make a high mutation score impossible
      causes raw oracle gaps to tend positive, \emph{covered} oracle gaps are
      both negative and positive.  They cluster around zero,
      indicating a kind of ``default'' balanced test effort.  However,
      many projects have much larger gaps, suggesting
      more effort should be spent in improving the ``lagging'' measure.
    \item Our manual review suggested that large positive covered oracle gaps
      often clearly indicated an actionable lack of important assertions.
\end{itemize}

Since both Figures \ref{fig:alllr} and \ref{fig:coveredlr} showed significant noise,
we also computed the variance of the oracle gap within projects as compared to overall variance. 
Average project variance was
{\averageprojvariance}, compared to the overall variance of {\averagevariance}.
Projects tend to have a higher degree of similarity with
regards to their oracle quality even across differently-covered files as
compared to the aggregate variance across all files (i.e, 
projects do, even within large variance, have a ``testing style'').

Finally, we note that a rank listing of files by mutation score may
likely be
very similar to a ranking by coverage.  While our data does not show
this for individual projects (we did not collect enough files per
project), we note that coverage and mutation score rankings of all
files have a Spearman correlation of 0.56.
In contrast, coverage and covered oracle gap have correlation of 0.53
and mutation score and covered oracle gap have a correlation of only
0.28.  Considering only covered mutation score lowers the correlation
further to 0.13.  Of course, end users will seldom wish to rank files
across many projects, but the relative relationship (where coverage
and mutation score provide fairly redundant information about testing,
but oracle gap provides an orthogonal ranking) seems likely to exist
within projects as well.  A rank listing by oracle gap will highlight
cases where mutation score may technically be ``good'' but lags
coverage by an unusual amount; this is never clear from mutation score
or coverage \emph{alone}.

\section{Detecting testing problems}
\label{sec:synthetic}

Our analysis of trends in Section~\ref{sec:javastudy} suggests that oracle gaps
\emph{may} be informative, especially in
projects with better coverage, and on covered code only.  However, it is a
purely descriptive study; being retroactive, it does not fully establish that
the oracle gap can pinpoint problems in a testing effort \emph{as they occur}.
This motivates our third research question: 
\vspace{-1em}
\begin{quote}
\item \textbf{RQ3}: Can oracle gap clarify testing \emph{problems}?
\end{quote}


To answer this question, we artificially induce ``problems'' in testedness by synthetically varying the coverage and oracle power
of test suites for 
well-tested Java files, and observe the corresponding behavior of the oracle
gap. 
We do this to confirm that the
(covered) oracle gap does, in general, ``advise'' the testers of these files to
focus on either coverage or oracle power, respectively, in the appropriate
direction.
We also
compare to pseudotesting~\cite{pseudotestingorig}, an alternative
test adequacy metric, to show that the oracle gap is a more effective framework
for test evaluation. 
We supplement this with a qualitative discussion of how the oracle gap could
have informed testing efforts in real-world code and changes. 


\noindent\textbf{Dataset.} These experiments require high quality test suites,
and significant computation time. We therefore selected 25 files (and their
tests) from 5 projects in our Java dataset with both high statement coverage and high mutation
score; 
The left side of Table~\ref{tab:syntheticstatsfull} summarizes this smaller corpus. 
We manually examined these files' unit tests to ensure they could be
manipulated as necessary.

\subsection{Synthetic Experiment}

The goal of this experiment is to artificially inject testedness problems into a
test effort for a file. 
We confirm, in general, that the covered oracle gap
detects these synthetic induced issues and provides the appropriate actionable advice.

\subsubsection{Setup and Analysis}
For each file, we programmatically find all related tests and \texttt{assert}
statements.
We generate synthetic test suites for these 25 well-tested Java files by
sampling from the starting pool of assertions and tests for each file
(effectively by commenting out some number of assertions and tests to produce a
smaller suite).  Our configurations consist of 0\%, 50\%, and 100\% of both
assertions and tests, in combination.  The intuition here is that we construct, at different levels, a range of test suites from high-coverage but low-quality, to low-coverage but high-quality, and high-coverage and high-quality (the original). For example, the \emph{50\% tests, 100\%
assertions} configuration consists of half the tests but all of those tests' assertions;
100\% tests, 0\% assertions includes all tests, but with all assert statements
removed. We exclude configurations with 0\% of the tests and more than 0\% of
assertions.  For any configuration involving 50\% of either assertions or tests, we
randomly sample five times, unless the number of assertions per test in a file
is less than 1.5 on average. We compute coverage and mutation analysis on the source code
files in question using the resulting test suites to generate both the regular
oracle gap and the covered oracle gap. 
The mutation analysis over this corpus for these experiments
took 20 CPU weeks over 19
AWS EC2 instances.

We exclude integration or system tests in these experiments, and unit tests that check for exceptional behavior.
We do this because we 
modify tests by removing 
\texttt{assert} statements; in
tests checking for exceptional behavior, removing the \texttt{assert} means that
thrown exceptions are not checked, leading the test to always (incorrectly) fail.
System and integration
tests are not as simply structured as unit tests, making it
difficult-to-impossible to programatically systematically vary test coverage/oracle strength.
The 0\% tests provide baselines, measuring adequacy of
the system/integration tests and unit tests for
exceptional behavior alone.

\subsubsection{Results}


  \begin{figure}%
    \centering
    \subfloat[100:100 difference: element config - 100:100 config\label{fig:gapdeltas_100}]{{\includegraphics[width=7cm]{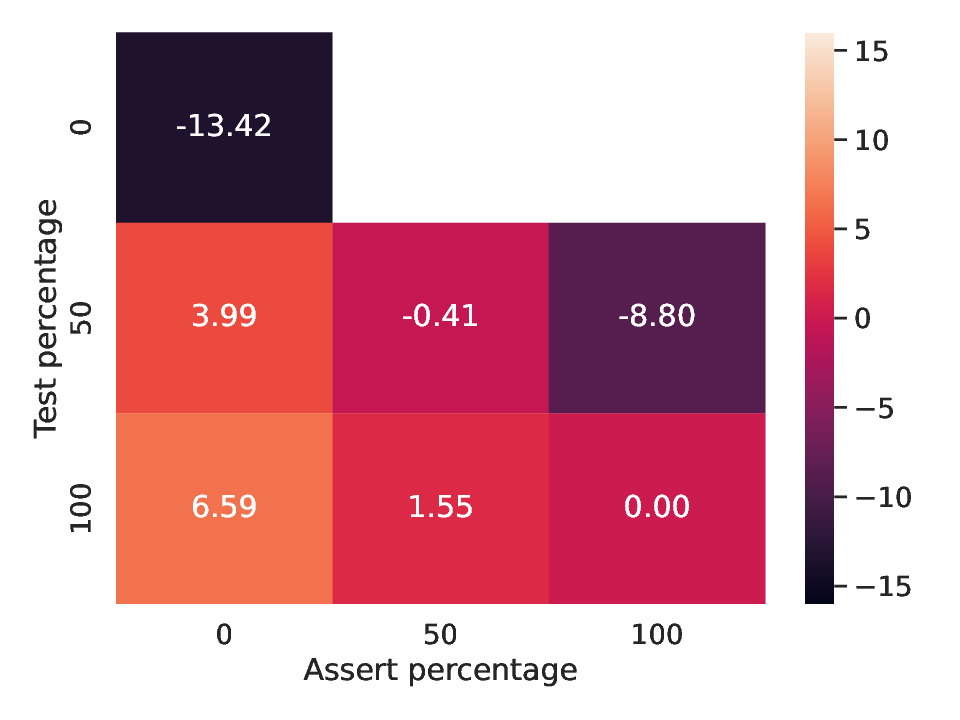} }}%
    \qquad
    \subfloat[Previous difference: element config - left of element (wrapping to row above if leftmost in current row) config\label{fig:gapdeltas_prev}]{{\includegraphics[width=7cm]{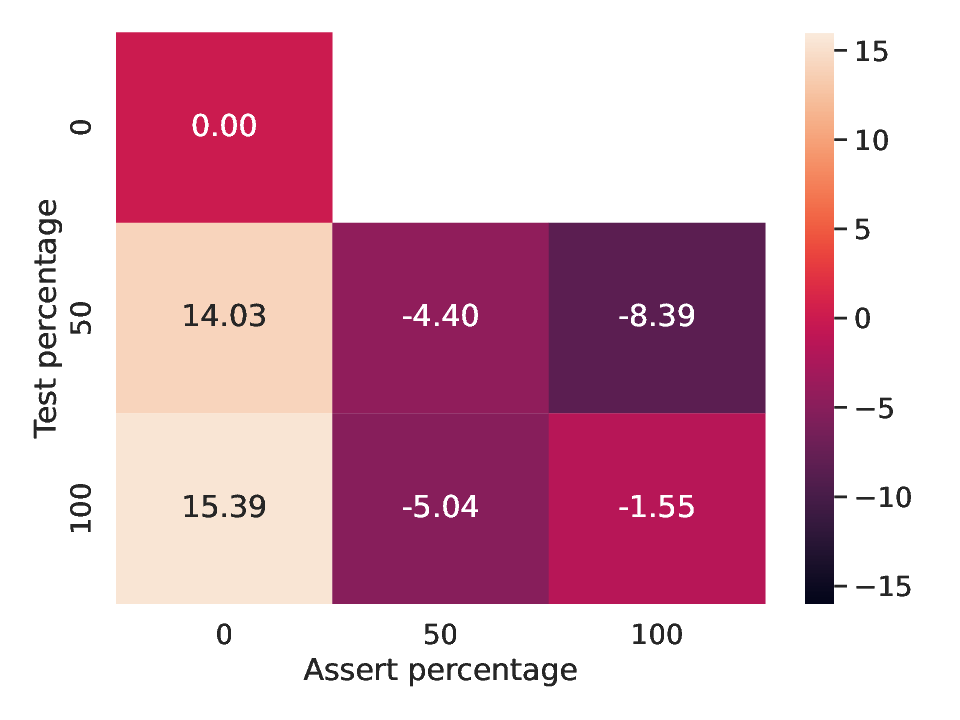} }}%
    \caption{\small Difference in mean covered oracle gap against either the full unit test suite, or the ``previous'' configuration. Covered oracle gaps fall as more asserts are added and rise fastest when tests with no asserts are added.}%
    \label{fig:gapdeltas}%
\end{figure}

Figure~\ref{fig:gapdeltas_100} shows
how the covered oracle gap changes with respect to the \emph{full} 
suites (100\% of tests and 100\% of asserts); Figure~\ref{fig:gapdeltas_prev}
shows how the covered oracle gap changes with respect to the \emph{previous}
test suite configuration
(the configuration to the
left). These differences are averaged over all configuration executions
on all 25 files.  To understand this matrix better, consider one of the files in our corpus, 
\texttt{ShardingService.java} (raw numbers for this file are available in our replication package).  At 0\% of
unit test cases (0\% assertions), the tests for this file show a covered oracle gap of
21.10\%. This means that the integration and exceptional behavior unit tests
are substantially biased in the direction of coverage over oracle strength. This
gap grows as more tests are added without their assert statements: 
100\% of the tests with 0\% of the asserts have a covered oracle gap of 40.63\%.

We also investigate what happens when a hypothetical tester added more asserts.
For example, 100\% of tests with 50\% of available assertions
consists of all the system and integration tests along with 100\% of the test cases in the unit test file with
50\% of their constituent assertions. Note that any configuration including 50\% of tests
or assertions are sampled to account for variance. 
When averaged across all five samples, the tests for our example file achieve a covered oracle gap
of 31.75\% at the 100\% of tests with 50\% of asserts, which is lower than the covered oracle gap of 40.53\% 
at the 100\% of tests with no assertions. Adding assert statement reduces
overall oracle gap.  In other words, when assertions are ``missing''
the oracle gap straightforwardly indicates a tester should 
add assertions/oracle power.

\begin{table*}
\caption{Synthetic experiment corpus, consisting of 25 well-tested Java files
  over 5 projects.  Right-hand columns show covered mutation score, gaps, and descartes pseudotesting scores. \label{tab:syntheticstatsfull}}
\centering
{\footnotesize
\begin{tabular}{llrrrr|rrrr}
\toprule
                          & \bf                                 & \bf Test &             &           &             & \bf Pseudo & \bf Covered    & \bf Covered & \bf Pseudo \\
\bf Project               & \bf Tested File                     & \bf LOC  & \bf Mutants & \bf Tests & \bf Asserts & \bf Score  & \bf Mut. Score & \bf Gap     & \bf Gap \\
\midrule
\parbox[t]{2mm}{\multirow{5}{*}{\rotatebox[origin=c]{35}{datasketches}}}
                          & DirectQuickSelectSketch              & 683     & 366         & 22       & 151          & 100\%    & 61.3\%     & 34.1\%       & -4.6    \\
                          & ReservoirItemsSketch                 & 458     & 604         & 14       & 85           & 96\%     & 68.5\%     & 30.2\%   & 2.7\%         \\
                          & Util                                 & 279     & 1133        & 17       & 74           & 93\%     & 73.4\%     & 25.0\%   & 5.4\%         \\
                          & BoundsOnBinomialProportions          & 104     & 702         & 4        & 12           & 100\%    & 92.0\%    & 8.5\%    & 0.5\%         \\
                          & MurmurHash3Adaptor                   & 297     & 741         & 17       & 53           & 74\%     & 50.7\%    & 48.0\%   & 24.7\%        \\
\midrule
\parbox[t]{2mm}{\multirow{5}{*}{\rotatebox[origin=c]{35}{apollo}}}
                          & StubClient                           & 242  & 44         & 17       & 24               & 88\%     & 66.7\%     & 26.1\%       & 4.8\%    \\
                          & JsonSerializerMiddlewares            & 70   & 10         & 5        & 9                & 100\% & 100\%     & -5.0\%   & -5.0\%     \\
                          & RuleRouter                           & 273  & 53         & 29       & 60               & 100\% & 91.4\%    & -18.9\%  & -27.5\%    \\
                          & Headers                              & 102  & 35         & 9        & 21               & 100\% & 83.9\%    & 16.1\%   & 0.0\%      \\
                          & ServiceImpl                          & 531  & 259        & 41       & 77               & 83\%  & 60.3\%    & 33.8\%   & 11.1\%     \\
\midrule
\parbox[t]{2mm}{\multirow{5}{*}{\rotatebox[origin=c]{35}{lmdbjava}}}
                          & CursorIterable                       & 271  & 51         & 21       & 47               & 100\%    & 82.5\%     & 12.2\%       & -5.3\% \\
                          & KeyRange                             & 233  & 35         & 19       & 41               & 100\% & 77.8\%    & 22.2\%   & 0.0\%      \\ 
                          & ByteBufferProxy                      & 118  & 187        & 7        & 11               & 87\% & 76.4\%     & 15.1\%   & 4.5\%      \\ 
                          & ResultCodeMapper                     & 124  & 32         & 5        & 7                & 100\% & 96.2\%    & 3.8\%    & 0.0\%      \\ 
                          & Txn                                  & 257  & 113        & 7        & 19               & 100\% & 79.1\%    & 19.6\%   & -1.3\%     \\ 
\midrule
\parbox[t]{2mm}{\multirow{5}{*}{\rotatebox[origin=c]{35}{github-api}}}
                          & GHGist                               & 100  & 51         & 3        & 45                & 75\%     & 76.3\%     & 20.3\%       & 21.6\%    \\
                          & GHLicense                            & 120  & 29         & 10       & 40                & 86\% & 50.0\%     & 44.2\%   & 8.2\%      \\
                          & GHCheckRunBuilder                    & 114  & 131        & 6        & 23                & 100\% & 97.9\%    & -2.9\%   & -5.0\%     \\
                          & GHWorkflow                           & 122  & 44         & 5        & 22                & 80\% & 80.0\%     & 9.1\%    & 9.1\%      \\
                          & GHBranchProtection                   & 87   & 14         & 6        & 18                & 60\% & 50.0\%     & 21.2\%   & 11.2\%     \\
\midrule
\parbox[t]{2mm}{\multirow{5}{*}{\rotatebox[origin=c]{35}{ss-elasticjob}}}
                          & ElasticJobExecutor                   & 205  & 69         & 8        & 20                & 100\%    & 69.6\%     & 21.5\%       & -8.9\%    \\
                          & ShardingService                      & 246  & 146        & 18       & 37                & 85\% & 71.0\%     & 29.0\%   & 15.0\%     \\ 
                          & AverageAllocationJobShardingStrategy & 56   & 88         & 6        & 6                 & 100\% & 88.8\%    & 11.3\%   & 0.1\%      \\ 
                          & AppConstraintEvaluator               & 146  & 127        & 7        & 15                & 100\% & 64.2\%    & 35.8\%  &  0.0\%      \\ 
                          & TaskContext                          & 92   & 110        & 13       & 27                & 100\% & 94.7\%    & 4.4\%   & -0.9\%      \\ 

\bottomrule
\end{tabular}}
\end{table*}

Figures \ref{fig:gapdeltas_100} and \ref{fig:gapdeltas_prev}, show that the
covered oracle gap always decreases as unit test assertions are added. Practically, this
means that when covered oracle gaps are strongly positive, one can directly add more
asserts to decrease the covered-oracle gap, and improve overall test
oracle power. Conversely, directly adding coverage with minimal asserts (for example 0-0
to 100-0), directly increases the covered oracle gap --- coverage itself
increases by construction, and the oracle gap precisely pinpoints the fact that
oracle power remains weak. 

While the average trends of oracle gaps are in line with our hypotheses,
the trend does not hold universally for all files. Specifically, while in all files
adding asserts increases mutation score, coverage is not always fixed as
assert statements are added.
This is because asserts sometimes contain invocations to source methods, increasing coverage
and oracle power simultaneously. In rare cases coverage from these asserts is greater than the corresponding increase in mutation
score.  That said, one way to think about the synthetic experiment is to note
that on average, \emph{following the advice of covered oracle gap}
tends towards producing the final test suites.  If we assume the final
suites for popular mature projects are often also high quality,
then the advice is generally good.

\subsection{Comparison with Pseudotesting}

Our observations so far demonstrate that the oracle gap is more actionable than
mutation score or coverage alone. We next compare it against pseudotesting,
which uses very coarse-grained mutants to more efficiently estimate mutation
score. This is done by replacing each source method body with a single line
returning some default value (e.g., replacing the body of a method that returns
a \texttt{String} with \texttt{return "";}) and checking
whether test behavior changes.  This fundamentally means that it
cannot identify cases where the lack of oracle power is nuanced; even
such large mutations as statement deletions are generally under its threshold.
%
We use the same corpus as above to ensure an apples-to-apples discussion; we
used \texttt{descartes}~\cite{descartes} (a PIT pseudotesting tool) to measure
each file's pseudotesting score.

The right-hand columns of Table~\ref{tab:syntheticstatsfull} show that for
these well-tested files, pseudotesting and mutation scores vary dramatically. We
suspect that this is because these are well-covered files from highly starred
Java projects, meaning that it's rare for methods to be completely untested. 
For such files, 
these results indicate that pseudotesting does not serve as a valid substitute
to computing the oracle gap, since the pseudotesting numbers suggest that the
test suites are more than adequate.

Pseudotesting and the oracle gap are complementary metrics.
Due to the significantly lower
compute cost required for pseudotesting, it may be useful early in a
testing effort, when entire methods are untested; however, when oracle
omissions are less blatant, it may offer little actionable information.

\subsection{Qualitative Discussion}

\begin{figure}
\begin{lstlisting}[language=Java, basicstyle=\footnotesize\ttfamily]
public boolean get(final T key, final T data, final SeekOp op) {
  if (SHOULD_CHECK) {
    requireNonNull(key);
    requireNonNull(op);
    checkNotClosed();
    txn.checkReady();
  }
  kv.keyIn(key);
  // ... elided happy path ...
\end{lstlisting}
\caption{\small The function in \texttt{Cursor.java} in the lmdbjava newly
  covered by a test added in commit \texttt{a781b}\label{fig:getfig}}
\end{figure}

Our injection of ``testedness problems'' into high-quality test
suites shows that the oracle gap would give the right ``advice'' to test
engineers.  But, the injection is by definition artificial.  We therefore
additionally manually examined recent real-world commits that significantly
changed tests on these projects to see how the oracle gap applies in current
development practice.  For many commits, coverage and mutation score change in
roughly equal proportion.  
%
%
However, here we highlight two (of several) interesting cases we found as
anecdotal examples of how the oracle gap could assess and inform current testing
efforts.

In \texttt{lmbdjava} project, commit
\texttt{a781b}
adds a new test to cover the \texttt{get} method from the
\texttt{Cursor} class,
shown in Figure~\ref{fig:getfig}.
As expected, both raw coverage and raw mutation score increase (from 88.6\% to
99.6\% and 41.1\% to 47.9\%, respectively).  The new test covers new code and as a result kills more mutants. However, the 
covered oracle gap of this file actually \emph{increases} from 47.5\% to 51.6\%, because the
test only checks the \emph{happy} execution path---none of the checks on lines
3--6 in Figure~\ref{fig:getfig} fail.  Although both coverage and mutation
score go up---substantiating that adding this new test was good---the change in
the oracle gap more precisely identifies the weaknesses that remain, and could
have motivated the developer to more closely examine the new tests. 


In \texttt{shardingsphere-elasticjob}, commit \texttt{8e20d} 
adds a series of new tests that check various failure cases in the
underlying source class, such as checking that jobs are paused when the computer
is shutdown or that a \texttt{JobFailureException} is appropriately thrown under exceptional cases. As
expected, coverage does not change much, going from 97.5\% to
97.6\%.  However, raw mutation score increases from 63.9\% to 72.2\%.
Covered oracle gap therefore decreases substantially
from 33.7\% to 25.4\%.  

These examples show how coverage, while informative, is limited on its own as
an adequacy metric.  Mutation score alone, however, does not fully supplant it
and, when looked at alone, may not be cost effective in terms of developer
time~\cite{GoogleMut}.  For \texttt{lmdbjava}, mutation score \emph{does}
increase with the new tests.  Given finite time and the fact that 100\% mutation
score is intractable, when should a developer spend more time looking at the remaining
unkilled mutants? On this commit, the oracle gap gives a much clearer
indication that the developer might fruitfully consider whether the new tests
are as strong as they could be.  Meanwhile, coverage does not increase
meaningfully in the \texttt{shardingsphere-elasticjob} example, but the
narrowing oracle gap shows that the new tests are adding significant strength to
the overall suite. In other cases a low mutation score
might, on its own, indicate a weak oracle, when in fact the problem is
entirely due to poor coverage.

\section{Oracle Gap Case Study}
\label{sec:crypto}

Having investigated the oracle gap in the large, we ask:
\begin{quote}
\textbf{RQ4}: What are the implications (and causes) of a small or
large, and positive or negative, oracle gap, across comparable real-world test
efforts? 
\end{quote}

Our Java studies show general trends and examples of oracle
gaps within projects, and so speak to the use of finding files with
actionable gaps in a project, but do not address ``higher level''
uses comparing testing ``styles.''  The diversity in kinds of code
makes drawing conclusions across different testing efforts
difficult.  We thus sought an example of high-criticality code of similar
functionality and complexity \emph{across projects}.

\subsection{Experimental Setup}

In particular, we examined the transaction validation code for several cryptocurrency
projects. 
Cryptocurrencies are essentially the sum of the operations of code executed by many independent nodes, especially nodes that mine cryptocurrency.  
The code validating blocks and transactions for blockchains
implementing cryptocurrencies is therefore of the utmost importance:
arguably, checking transaction
correctness is the \emph{raison d'être} of any blockchain.

For Bitcoin Core~\cite{nakamoto2008bitcoin}, we chose the core {\tt tx\_verify.cpp} file, which verifies all
transactions. We also performed mutation analysis
of transaction-verification-related code for three other high market
cap cryptocurrencies: \texttt{ethereum}, \texttt{avalanche}, and
\texttt{algorand}.  Our selection was influenced by both the market cap of each
project and the availability of code coverage reports for the projects. We
identified candidate files roughly comparable to 
Bitcoin's {\tt tx\_verify.cpp} by searching for keywords like \texttt{transaction},
\texttt{verify}, \texttt{sign}, and \texttt{validate}. We manually inspected
functions and test coverage for these functions (where applicable) to identify
1--2 files focusing on this similar core functionality per
project to mutate.  We ran these mutants against each project's default test suite (determined by
consulting READMEs and build/CI documentation) using \um.
The projects are written in C++ and Go. 

\subsection{Results}

\begin{table*}[ht]
\caption{Covered Mutation Scores and Oracle Gaps For Selected Files}
\vspace{2mm}
\centering
\begin{tabular}{llccccr}
\toprule
\bf \mr{2}{Project}             & \bf \mr{2}{File path}                         & \mc{1}{c}{\bf File}       & \mc{1}{c}{\bf Mutation}  & \mc{1}{c}{\bf Covered} & \mc{1}{c}{\bf Raw}     & \mc{1}{c}{\bf Covered}   \\
\bf                             & \bf
                                                                                &
                                                                                  \mc{1}{c}{\bf coverage}   & \mc{1}{c}{\bf score}     & \mc{1}{c}{\bf mutation}            & \mc{1}{c}{\bf oracle}            & \mc{1}{c}{\bf oracle}  \\
  \bf                             & \bf
                                                                                &
                                                                                  \bf  &\bf    &   \mc{1}{c}{\bf score}           & \mc{1}{c}{\bf gap}     & \mc{1}{c}{\bf gap} \\
\midrule
bitcoin                         & src/consensus/tx\_verify.cpp                  & 98.7\%                    & 75.8\%                   & 83.1\%                          & 22.9\%                        & 15.5\%                   \\
\cmidrule{2-7}
\mr{2}{go-ethereum}             & core/block\_validator.go                      & 81.0\%                    & 66.7\%                   & 78.0\%                          & 14.3\%                        & 3.0\%          \\
                                & signer/fourbyte/validation.go                 & 60.0\%                    & 42.2\%                   & 76.4\%                          & 17.8\%                        & -16.4\%                   \\
\cmidrule{2-7}
avalanchego                     & vms/platformvm/add\_subnet\_validator\_tx.go  & 79.1\%                    & 44.3\%                   & 67.1\%                          & 34.8\%                        & 12.0\%                   \\
\cmidrule{2-7}
go-algorand                     & data/transactions/logic/eval.go               & 90.0\%                    & 99.7\%                   & 99.7\%                          & -9.7\%                        & -9.7\%                   \\
\bottomrule
\end{tabular}
\label{tab:comparison}
\end{table*}

Table~\ref{tab:comparison} shows the range of raw and covered oracle
gaps for code in these cryptocurrencies.  
Our framework  exposes key
differences within otherwise superficially (by coverage or mutation score)
similar testing efforts on comparable code.  To return to the example from the
introduction, Bitcoin and go-algorand both have what would be usually
considered ``good'' coverage, at $>$ 90\%.  And the 75\% and 99\%
mutation scores are, respectively, solid and remarkable.  In
isolation, however, the numbers either suggest ``Bitcoin core is
somewhat better tested'' (coverage) or ``Algorand is much better
tested'' (mutation score).  Using both numbers, however, we can see
that these projects have chosen different trade-offs in testing.  The
Bitcoin core tests emphasize covering all code, and in fact do cover
all but extremely obscure (and possibly impossible to encounter in practice) behavior.  Meanwhile, Algorand
leaves more code uncovered, including code that appears to correspond
to obscure but still possible conditions.  However, Bitcoin
core has taken much less pain to construct strong oracles.  In
practice, 75\% is generally considered a very good mutation
score for real-world code, but the 99\%+ mutation coverage of
Algorand reflects an even greater attention to ensuring every
behavior's impact is checked, and moreover suggests that Algorand
chose to cover almost all code with serious semantic impact.
It is likely that to some extent Bitcoin sacrificed some
oracle-improvement effort to focus on coverage, which is
closely monitored by the project, while efforts to perform frequent mutation
analysis on Bitcoin Core are only now beginning.

Similarly, Ethereum's {\tt block\_validator.go} would benefit from stronger oracles. The tests for
{\tt validation.go} simply fail to cover a large portion of the file,
though do a good job of checking the covered code. 
For avalanchego, the covered oracle gap is what we could
consider ``normal'' reflecting a somewhat weaker oracle combined with
reasonable code coverage.   Note that by mutation score, {\tt
  validation.go} from Ethereum and {\tt add\_subnet\_validator\_tx.go}
appear to be very similar: using the conventional ``best
practice'' of applying mutation testing alone as an adequacy measure when
possible, these files would both seem to be poorly tested.  The
mutation score would not indicate that in one case, the failure is
mostly due to poor coverage (a large negative oracle gap) and in the
other case, the result is due to an oracle that significantly lags
coverage. As with Bitcoin and Algorand, 
considering the gap tells a much fuller story about the testing efforts, 
and the best mitigations for weaknesses. Scores for all projects show the importance of usually
focusing on \emph{covered} oracle gap for overall understanding: raw oracle gaps are all
positive, since at project level limited coverage dominates the factors
leading to unkilled mutants.

\subsection{Bitcoin and Fuzzing: Acting on the Gap}

Bitcoin Core includes a complex, well-designed, set of fuzzing tests that are
run on OSS-Fuzz~\cite{icseseip22}.  
Two particularly relevant fuzz targets in
{\tt tx\_verify.cpp} are ({\tt
  process\_message\_tx} and {\tt coins\_view}); we used these to explore qualitatively
how oracle gap is particularly informative for highly automated
testing methods.

 Fuzzing {\tt tx\_verify.cpp} does not affect code
coverage either way: fuzz test coverage is approximately as high as for
functional tests, though with minor changes in exact code covered.
However, overall, fuzzing could detect just under 12\%
of all the generated mutants, resulting in very large positive
oracle gaps. 
Fuzzing adds only two unique mutant kills beyond those that Bitcoin Core's
existing functional tests can find. 
%
%

So: why fuzz?  
The answer is that fuzzing
uncovers subtle bugs that functional tests designed by humans will
almost never detect,
e.g. \url{https://github.com/bitcoin/bitcoin/issues/22450}.  Neither fuzzing nor functional/unit tests
replace one another. Consider two
mutants detected by {\tt coins\_view} fuzzing alone. 
These correspond to two (hypothetical) bugs not detectable by any other means; in the
real world, if one such bug is exploitable, detecting it may ``pay
for'' all the fuzzing effort, and there will seldom be just one such
bug (see an approximate list of fuzzer-detected, fixed bugs in Bitcoin
Core at \url{https://github.com/bitcoin/bitcoin/issues?q=is\%3Aissue+fuzz+is\%3Aclosed+label\%3ABug}).

While the oracle gap does not show fuzzing is useless, it \emph{does} likely point to the most effective way to improve the fuzzing: manual, expert developer effort to improve the
\emph{oracles} used by existing fuzz targets, or efforts to craft custom,
more restricted, fuzz targets with stronger oracles. Building fuzz harnesses with complex correctness checks is hard, of
course; the functional tests know exactly what inputs are being
provided to APIs, and can check for expected behavior.  Trying to
inject this kind of check into fuzz harnesses ranges from non-trivial
to impossible.  When
applicable, more generic, ``mathematical'' constraints such as are
used in property-based testing~\cite{ClaessenH00} can help, but these
are often hard to devise.
The most promising route to solve these problems may be
to manually add high-quality invariants and assertions to source, which
can be executed by
both functional and fuzz tests. At
present, the Bitcoin Core code has about 1,800 {\tt assert}
statements, scattered among  ~180KLOC of C and C++.  The resulting ratio
of about one assertion per 100 lines of code is not terrible, but is certainly
at the lower limit of acceptable. 
Given that Bitcoin Core defines at least 4,000 functions, the code fails 
to meet the NASA/JPL proposal of an average of two
assertions per function~\cite{holzmann2006power}.  

\section{Threats to Validity}

The primary threats to validity of which we are aware are ones common
to all studies using mutation testing, e.g., equivalent mutants;
however, as our interests are in correlations between coverage and
mutation score, rather than absolute mutation scores, we do not believe
these are significant.  A second potential threat is that the cryptocurrency
study only examined a small number of projects and a particular
functionality, so may not generalize.  The threat of a focus on Java/Maven code in most of our
results is however mitigated by similar findings over C++ and Go
code in the cryptocurrency projects.  Our replication package
(\url{https://figshare.com/s/06cbb520b30751ebb1b2}) provides full
details on all experiments and methods.

\section{Related Work}

While
there is foundational work on the oracle/test
distinction~\cite{StaatsOracle} at a theoretical level, most previous work assumes a basic framework of  trying to determine
correlation of mutation testing or coverage alone with fault
detection and/or coverage with mutation score~\cite{papadakis2018mutation,PapadakisStudy,ThierryStudy}.  In
such work, the oracle/test distinction is not considered in the
light of whether code is ``more executed'' or ``more checked.''  The
most similar approach, which does not involve mutants, is the notion
of \emph{checked coverage}~\cite{ZellerCheckedCov}, which, however, still results in a
single score, as does pseudotesting~\cite{pseudotestingorig,
  pseudotestingstudy, descartes}.  Checked coverage is
inherently unable to identify \emph{buggy} assertions that do not in practice
detect mutants, if there is any dynamic flow
from coverage to the faulty assertion.  The study of coverage gaps (in
a different sense than ours) of Hossain et al.~\cite{GapStruct}
extends checked coverage to further coverage criteria, and provides
\emph{recommendations} (at the individual statement level only, unlike
our approach) for code that may not be checked, but does not
escape this fundamental limitation.  Their study does, however, show
that gaps of the type we identify more completely often cause
faults to go undetected.  

Just et al. \cite{JustMutationFault} found that 73\% of real-world faults can be associated with common mutation
operators. Beller et al. \cite{BellerFacebookMutation} examined how to feasibly
implement mutation testing at Facebook. These studies support the notion that
lagging mutation score indicates lagging fault \emph{detection}
capability despite high coverage, as also suggested by work
by Sina et al. on whether automatically generated unit tests help find bugs~\cite{DoGenerated}.
Smith et al. \cite{SmithCoverageMutation} briefly discussed the
relationship between coverage and mutation score, but 
examined only two open source projects. Li et al. \cite{LiCoverageMutation} also examined when it is best to use various testing methods, including mutation testing. Recent industrial uses of
mutants~\cite{PetrovicMutationGoogle,BellerFacebookMutation} do not
include gap analysis, but informally hint that large-scale users may
be amenable to the type of analysis we provide, given how they use
mutants to pinpoint test issues now.  Finally, it is interesting to consider that
large positive covered oracle gap can be alternatively thought of as
the presence of many \emph{stubborn} mutants in a file~\cite{papadakis2018mutant}.

\section{Conclusion and Implications}

Current approaches to measuring test adequacy focus on either simply
reporting structural coverage, or on reporting a complex mix of oracle
power and structural coverage, in the form of mutation score.  Neither
approach yields easily practicable advice on the degree to which a test's
coverage and oracle efforts are ``in balance.''  We propose the
\emph{oracle gap} or
\emph{the difference between source code coverage and mutation score} as a new
metric to help researchers and practitioners improve their
understanding of test adequacy.  We show that in practice the oracle
gap varies widely in real code, and connect that variance to real
differences in testing.  Our core immediate \emph{use} for the
oracle gap, identifying cases of missing or \emph{faulty} assertions
 \emph{in important code} is
supported by showing that removal of assertions creates substantially
larger gaps.

This way of looking at test adequacy has implications for both testing researchers and
practitioners. For researchers, while there exists the expected correlation 
between coverage and mutation score, the
relationship is subtle: mutation score is not a ``refined''
coverage score, and reporting one number without the context of
the other paints a partial picture, especially for automatic
test generators.  There is already strong impetus for research in improving or
constructing oracles for existing or generated tests; our results show that the
recent growth and impressiveness of fuzz testing efforts provides
further motivation for examination of gaps.

This speaks to the implications of our reasoning for practitioners.  
Covering more code probably \emph{is} good, but there is a balance in how finite testing
effort should be allocated.
Practitioners can look at the
oracle gap from time to time in their test efforts to understand where their
efforts should next be allocated.  Companies 
are already using mutants of covered lines that are not killed to
improve testing.  In addition to localized advice, oracle gap might give
visibility into what might be called \emph{oracle debt} by analogy with technical
debt: methods, files, or entire submodules where policies about coverage have yielded testing whose coverage
exceeds its ability to probe for faults, or where buggy assertions
make effort expended fail to pay off.

\small{A portion of this work was supported by the National Science
Foundation under awards CCF-2129446 and CCF-2129388.}

\bibliographystyle{plain}
\bibliography{bibliography}

\end{document}